# Compact and scalable polarimetric self-coherent receiver using dielectric metasurface

**GO SOMA,**[1,4] **YOSHIRO NOMOTO,**[2] **TOSHIMASA UMEZAWA,**[3] **YUKI YOSHIDA,**[3] **YOSHIAKI NAKANO,**[1] **AND TAKUO TANEMURA**[1,5]

[1]*School of Engineering, The University of Tokyo, 7-3-1 Hongo, Bunkyo-ku, Tokyo, Japan*
[2]*Central Research Laboratory, Hamamatsu Photonics K.K. 5000 Hirakuchi, Hamakita-ku, Hamamatsu City, Shizuoka, Japan*
[3]*National Institute of Information and Communications Technologies (NICT), 4-2-1 Nukui-Kitamachi, Koganei, Tokyo, Japan*
[4]*soma@hotaka.t.u-tokyo.ac.jp,* [5]*tanemura@ee.t.u-tokyo.ac.jp*

**Abstract:** The polarimetric self-coherent system using a direct-detection-based Stokes-vector receiver (SVR) is a promising technology to meet both the cost and capacity requirements of the short-reach optical interconnects. However, conventional SVRs require a number of optical components to detect the state of polarization at high speed, resulting in substantially more complicated receiver configurations compared with the current intensity-modulation-direct-detection (IMDD) counterparts. Here, we demonstrate a simple and compact polarimetric self-coherent receiver based on a thin dielectric metasurface and a photodetector array (PDA). With a single 1.05-µm-thick metasurface device fabricated on a compact silicon-on-quartz chip, we implement functionalities of all the necessary passive components: a 1×3 splitter, three polarization beam splitters with different polarization bases, and six focusing lenses. Combined with a high-speed PDA, we demonstrate self-coherent transmission of 20-GBd 16-ary quadrature amplitude modulation (16QAM) and 50-GBd quadrature phase-shift keying (QPSK) signals over a 25-km single-mode fiber. Owing to the surface-normal configuration, it can easily be scaled to receive spatially multiplexed channels from a multicore fiber or a fiber bundle, enabling compact and low-cost receiver modules for the future highly parallelized self-coherent systems.

## 1. Introduction

Rapid spread of cloud computing, high-vision video streaming, and 5G mobile services has led to a steady increase in information traffic in the datacenter interconnects and access networks [1]. While intensity-modulation direct-detection (IMDD) formats such as 4-level pulse amplitude modulation (PAM4) are employed in the current short-reach optical links, scaling these IMDD-based transceivers beyond Tb/s is challenging due to the limited spectral efficiency and severe signal distortion caused by the chromatic dispersion of fibers. On the other hand, the digital coherent systems used in metro and long-haul networks can easily expand the capacity by utilizing the full four-dimensional signal space of light and complete compensation of linear impairments through digital signal processing (DSP). However, substantially higher cost, complexity, and power consumption of coherent transceivers have hindered their deployment in short-reach optical interconnects and access networks.

To address these issues, the self-coherent transmission scheme has emerged as a promising approach that bridges the gap between the conventional IMDD and coherent systems [2-7]. In this scheme, a continuous-wave (CW) tone is transmitted together with a high-capacity coherent signal, which are mixed at a direct-detection-based receiver to recover the complex optical field of the signal. Unlike the full coherent systems, this scheme eliminates the need for a local oscillator (LO) laser at the receiver side as well as the stringent requirement of using wavelength-tuned narrow-linewidth laser sources, suggesting that substantially low-cost broad-linewidth uncooled lasers can be used [4]. In addition, since the impacts of laser phase noise and frequency offsets are mitigated, the computational cost of DSP can be reduced significantly [7, 8]. The self-coherent systems thus enable low-cost, low-power-consumption, yet high-capacity data transmission, required in the future datacenter interconnects and access networks.

Among several variations of implementing self-coherent systems, the polarimetric scheme using a Stokes-vector receiver (SVR) [9-11] has an advantage in terms of simplicity. In this scheme, the coherent signal is transmitted on a single polarization state, together with a CW tone on the orthogonal polarization state. By retrieving the Stokes parameters $\mathbf{S} = [S_1, S_2, S_3]$ at the receiver side, the in-phase-and-quadrature (IQ) signal is demodulated through the DSP after compensating for the effects of polarization rotation, chromatic dispersion, and other signal distortions. To date, a number of high-speed polarimetric self-coherent transmission experiments have been reported, where the SVRs were implemented using off-the-shelf discrete components [3, 10-12]. Toward practical use, integrated waveguide-based SVRs were also realized on Si [13, 14] and InP [15-18]. More recently, surface-normal SVRs were demonstrated using nanophotonic circuits [19, 20] and liquid crystal gratings [21] with external photodetectors (PDs). Compared with the conventional low-cost IMDD receivers, however, these devices still suffer from a large fiber-to-chip coupling loss and/or need for external lenses to focus light to PDs.

In this paper, we demonstrate high-speed polarimetric self-coherent signal detection using a compact surface-normal SVR, composed of a metasurface-based polarization-sorting device and a high-speed two-dimensional photodetector array (2D-PDA). A metasurface is a two-dimensional array of subwavelength structures that can locally change the intensity, phase, and polarization of input light [22]. Unlike the previous works on metasurface-based polarimeters for imaging and sensing applications [23-27], our device enables efficient coupling of a self-coherent optical signal from a single-mode fiber (SMF) and lens-less focusing to six high-speed PDs. More specifically, by superimposing three types of meta-atom arrays, it implements the functionalities of all the necessary passive components, namely a 1×3 splitter, three polarization beam splitters (PBSs) with different polarization bases, and six lenses, inside a single ultrathin device. Combined with an InP/InGaAs-based 2D-PDA chip, we demonstrate penalty-free transmission of polarimetric self-coherent signals over a 25-km SMF in various formats such as 20-GBd 16-ary quadrature amplitude modulation (16QAM) and 50-GBd quadrature phase-shift keying (QPSK). Owing to the surface-normal configuration with the embedded focusing functionality, highly efficient lens-free coupling to the 2D-PDA is achieved. The demonstrated SVR, therefore, has a comparable complexity as a conventional low-cost IMDD receiver that fits in a compact receiver optical subassembly (ROSA). Moreover, it can readily be extended to receive spatially multiplexed channels from a multicore fiber (MCF) or a fiber bundle, which are expected in the future >Tb/s highly parallelized optical interconnects [28-31].

## 2. Device concept

The schematic of the proposed surface-normal SVR is illustrated in Fig. 1(a). The light from an SMF is incident to a thin metasurface-based polarization-sorting device, which is designed to provide the same functionality as a conventional polarimeter shown in the inset. Namely, it splits the light into three paths, resolves each of them to the orthogonal components in three different polarization bases, and focuses them to six PDs integrated on a 2D-PDA chip. Unlike previously demonstrated metasurface-based polarimeters [23-27], our proposed SVR implements the 1×3 splitter and six metalenses as well to enable direct coupling from an SMF to a high-speed 2D-PDA. As a result, the entire device can fit inside a compact ROSA module, comparable to the current IMDD receivers. Moreover, owing to the surface-normal configuration, this scheme can easily be scaled to receive multiple spatial channels without increasing the number of components by simply replacing the input SMF to a MCF or a fiber bundle and using a larger-scale PDA [32] as shown in Fig. 1(b).

To enable three operations in parallel using a single metasurface layer, we adopt the spatial multiplexing method [33, 34]; three independently designed meta-atom arrays are superimposed as shown by MA1 (red), MA2 (blue), and MA3 (green) in Fig. 1(c). The phase profile $\varphi(x, y)$ of MA1 is designed to focus the $x$-polarized component of light to $PD_x$ and the $y$-polarized component to $PD_y$ at the focal plane as shown in the inset. Similarly, MA2 and MA3 function as PBSs with embedded metalenses for the ±45° polarization basis (a/b) and the right/left-handed circular (RHC/LHC) polarization basis (r/l), respectively, and focus respective components to $PD_{a,b}$ and $PD_{r,l}$. The Stokes vector $\mathbf{S} \equiv (S_1, S_2, S_3)^T$ can then be derived by taking the difference of the photocurrent signals as $S_1 = I_x - I_y$, $S_2 = I_a - I_b$, and $S_3 = I_r - I_l$, where $I_p$ is the photocurrent at $PD_p$. We should note that this scheme with three balanced PDs without polarizers offers the maximum receiver sensitivity among various SVR configurations [35] and is advantageous compared with the previous demonstrations that employ a non-optimal polarization basis [19-21].

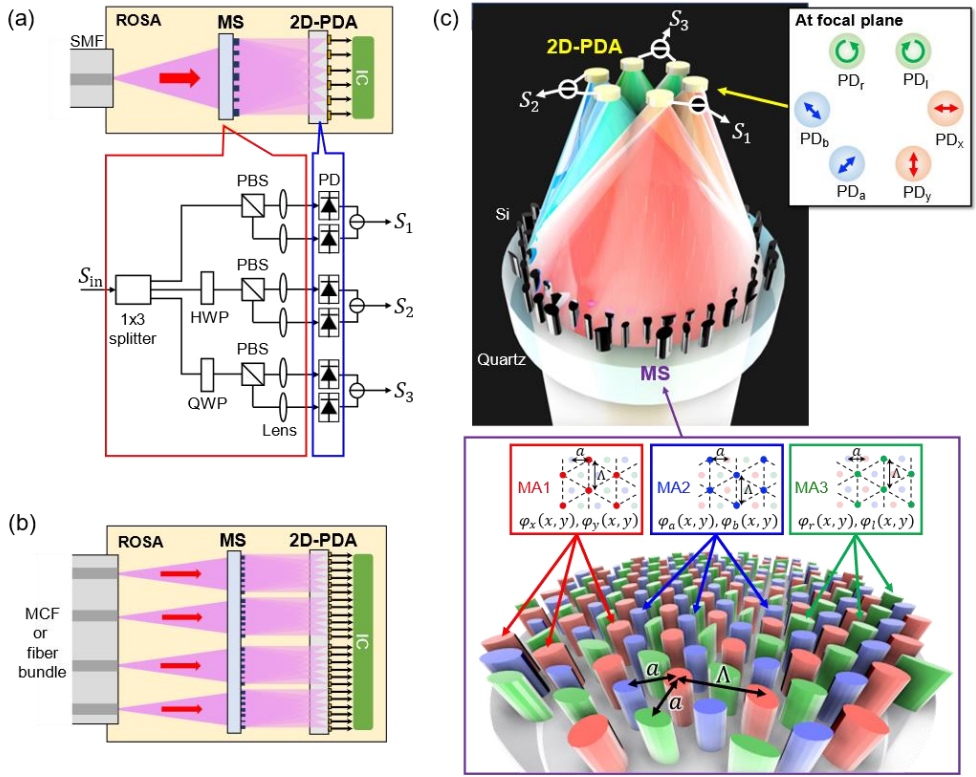

Fig. 1. Surface-normal SVR based on superimposed meta-atom arrays. (a) Schematic illustration of the receiver module. A single metasurface device implements all the necessary passive optical components of the equivalent circuit as shown in the inset. MS: metasurface. PDA: photodetector array. IC: integrated circuit. HWP: half-wave plate. QWP: quarter-wave plate. PBS: polarization beam splitter. (b) Scalable configuration to receive multiple input channels from a MCF or fiber bundle. (c) Functionality and configuration of the designed metasurface. The incident light from the SMF is split into three paths and focused to six PDs located at different positions according to the input state of polarization. The superimposed meta-atom arrays (MA1, 2, and 3) operate as PBSs and metalenses for $x/y$ linear, ±45° linear, and RHC/LHC polarization bases, respectively.

## 3. Metasurface design and fabrication

As the dielectric metasurface, we employ 1050-nm-high elliptical Si nanoposts on a quartz layer. The phase of the transmitted light and its polarization dependence can be controlled by changing the lengths of two principal axes $(D_u, D_v)$ and the in-plane rotation angle $\theta$ of each nanopost as defined in Fig. 2(a) [22]. Here, in each meta-atom array, MA1-3, we adopt the triangular lattice with a sub-wavelength lattice constant of $\Lambda = 700\sqrt{3}$ nm, so that the non-zero-order diffraction is prohibited. Then, three meta-atom arrays are superimposed by shifting their positions by $a = 700$ nm to form the overall metasurface, as shown in Fig. 1(c).

First, we set $\theta$ to 0 and simulate the transmission characteristics of uniform nanopost array for the *x*- and *y*-polarized light at a wavelength of 1550 nm by the rigorous coupled-wave analysis (RCWA) method [36]. From the simulated results, we first derive $t_u(D_u, D_v)$ and $t_v(D_u, D_v)$, which denote the complex transmittance for the *x*- and *y*-polarized light as a function of $D_u$ and $D_v$. Then, we derive the required $(D_u, D_v)$ that provides a phase shift of $(\varphi_u, \varphi_v)$ for each polarization component. The results are plotted in Fig. 2(b) (see Section S1 of Supplement 1 for details). The amplitude of transmittance for each case is also shown in Fig. 2(c). We can confirm that by setting the dimensions of the ellipse appropriately, arbitrary phase shifts for *x*- and *y*-polarized components can be achieved with high transmittance.

By rotating the elliptical nanoposts by $\theta$ as shown in Fig. 2(a), such birefringence can be applied to any linear polarization basis oriented at an arbitrary angle [37]. We should note that the phase shifts and amplitudes of transmission are nearly insensitive to $\theta$ [22] and similar results as shown in Fig. 2(b) and 2(c) are obtained for all $\theta$. This is because the light is strongly confined inside each Si nanopost, so that the optical coupling among neighboring meta-atoms has only minor influence on the transmission.

We can also provide arbitrary phase shifts to orthogonal circular-polarization states by using the geometric phase shift of meta-atoms [38]. First, we judiciously select $D_u$ and $D_v$ to satisfy $\varphi_v = \varphi_u + \pi$, so that each nanopost operates as a half-wave plate. In this case, input RHC and LHC states are converted to LHC and RHC, respectively. In addition, their phases after transmission are written as $(\varphi_r, \varphi_l) = (\varphi_u + 2\theta, \varphi_u - 2\theta)$ (see Section S2 of Supplement 1 for the derivation). Therefore, $D_u$ and $D_v$ of each nanopost are selected to obtain desired $\varphi_u$ $(=(\varphi_r + \varphi_l)/2)$ while satisfying the condition $\varphi_v = \varphi_u + \pi$. The angle $\theta$ is also determined to be $(\varphi_r - \varphi_l)/4$.

To realize the function of a metalens, each meta-atom array needs to impart a spatially dependent phase profile given as [39]

$$\varphi(x, y) = -\frac{2\pi}{\lambda}(\sqrt{(x - x_0)^2 + (y - y_0)^2 + f^2} - f), \qquad (1)$$

where $(x_0, y_0)$ is the in-plane position of the focal point, $f$ is the focal length, and $\lambda$ is the operating wavelength. In this work, we set $\lambda = 1550$ nm, $f = 10$ mm, and the diameter of the entire metasurface area to be 2 mm, corresponding to the numerical aperture (NA) of ~0.10. The six focal points are arranged on a regular hexagon with a spacing of 60 µm, which are matched to the positions of the high-speed 2D-PDA used in our self-coherent experiments. Under these conditions, the phase profiles required for MA1, 2, and 3 are determined as shown in Fig. 2(d). Note that a rather large (2 mm) metasurface is used in this work due to the limitation in reducing the focal length $f$ in the current optical setup. In a fully packaged module as shown in Fig. 1(a), we can readily shrink the entire area of the metasurface to a few tens of micrometers by reducing $f$ and designing the geometrical parameters of each nanopost to satisfy the required phase profiles given by Eq. (1).

The designed metasurface was fabricated using a silicon-on-quartz (SOQ) substrate with a 1050-nm-thick Si layer. The nanopost patterns were defined by electron-beam lithography with ZEP520A resist. Then, the patterns were transferred to the Si layer by inductively-coupled-plasma reactive-ion etching (ICP-RIE) using $SF_6$, $C_4F_8$, and $O_2$. An optical microscope image and scanning electron microscopy (SEM) images of the fabricated metasurface are shown in Fig. 2(e)-(g).

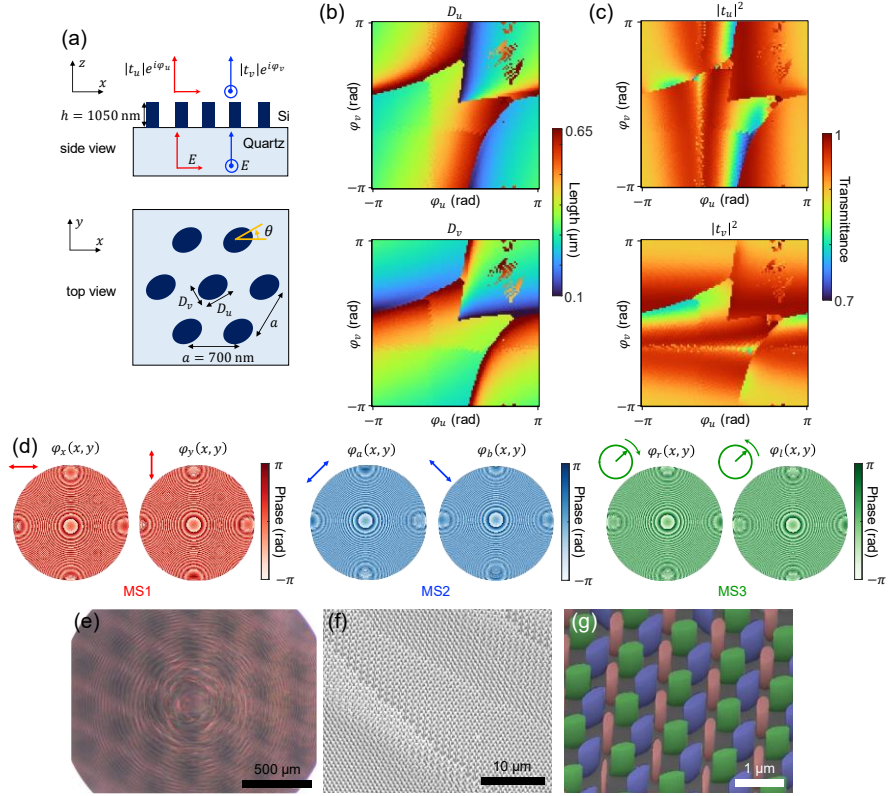

Fig. 2. Metasurface design and fabrication. (a) Schematic of a periodic array of Si nano-posts placed at the vertices of a triangular lattice with a lattice constant $a$ of 700 nm. The transmission of $x$- and $y$-polarized light is simulated for various axes lengths $(D_u, D_v)$ of the elliptical posts. (b) Required $(D_u, D_v)$ to obtain phase shifts $(\varphi_u, \varphi_v)$ for $x$- and $y$-polarized light. For ease of fabrication, the ranges of $D_u$ and $D_v$ are limited from 100 nm to 650 nm. (c) The amplitude of transmittance for each case in (b) as a function of $(\varphi_u, \varphi_v)$. (d) Required phase profiles for MA1, 2, and 3. (e) Optical microscope image and (f, g) SEM images of the fabricated device. In (g), the image is false-colored to distinguish MA1, 2, and 3.

## 4. Static characterization of the fabricated metasurface

We first characterized the fabricated metasurface by observing the intensity distribution at the focal plane for various input states of polarization (SOPs). The experimental setup is shown in Fig. 3(a). A CW light with a wavelength of 1550 nm was incident to the metasurface. The SOP was modified by rotating a half-wave plate (HWP) and a quarter-wave plate (QWP). The image at the focal plane was magnified at 50 times by a 4-f lens system and captured by an InGaAs camera. From the detected intensity values at the six focal positions, the Stokes vector was retrieved as described in Section 2. To enable quantitative measurement of the focused power, we inserted a flip mirror and detected the total power by a bucket PD after spatially filtering the focused beam at each target position using an iris.

Figure 3(b) shows the observed intensity distributions when the input Stokes vector is set to $(\pm 1, 0, 0)$, $(0, \pm 1, 0)$, and $(0, 0, \pm 1)$. We can confirm that the incident light is focused to the six well-defined points by transmitting through the metasurface. Moreover, its intensity distribution changes with the SOP; $x/y$ linear, $\pm 45°$ linear, and RHC/LHC components of light are focused to the designed positions as expected. Figure 3(c) shows the retrieved Stokes vectors on the Poincaré sphere. The average error $\langle|\Delta\mathbf{S}|\rangle$ is as small as 0.028. Figure 3(d) shows the measured focusing efficiencies to the six positions. Subtracting the 4.8-dB intrinsic loss due to the 1×3 splitter [see Fig. 1(a)], the excess loss is around 6.1 dB, whereas the crosstalk to the orthogonal PD position is suppressed by 13-20 dB. While this excess loss is already comparable to the coupling and propagation losses of the previously reported waveguide-based SVRs [13-18], we expect further improvement by applying anti-reflection coating at the silica surface, improving the fabrication processes to minimize the errors, and by adopting advanced algorithms in designing the metasurface that take into account the nonzero interactions between adjacent meta-atoms [40, 41].

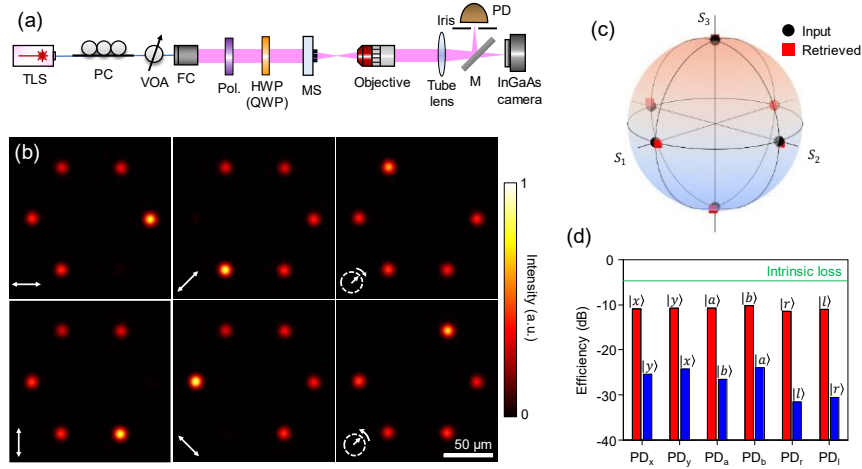

Fig. 3. Experimental characterization of the fabricated metasurface. (a) Schematic of the optical setup. The flip mirror is used to switch between capturing the intensity distribution at the focal plane and measuring the power of each focused beam. TLS: tunable laser source. PC: polarization controller. VOA: variable optical attenuator. FC: fiber collimator. Pol.: polarizer. HWP: half-wave plate. QWP: quarter-wave plate. MS: metasurface. M: flip mirror. PD: photodetector. (b) Measured intensity distributions at the focal plane for different input SOPs. (c) Retrieved and input Stokes vectors on the Poincaré sphere. (d) Measured focusing efficiency to each PD. The intrinsic loss due to splitting into three paths is shown by a green line. The input polarization is labeled on the top of each bar.

## 5.  Self-coherent signal transmission experiment

We then performed the polarimetric self-coherent signal transmission experiment using the fabricated metasurface. The experimental setup is shown in Fig. 4. We employed a 19-pixel 2D-PDA with InP/InGaAs-based p-i-n structure [42], from which six PDs were used as shown in Fig. 4(b). Each PD had a diameter of 30 µm, the measured bandwidth above 10 GHz, and the responsivity of 0.3 A/W. The 2D-PDA chip was packaged with the radio-frequency (RF) coaxial connectors connected to each PD. The 2D-PDA was placed at the focal distance of 10 mm from the metasurface as shown in Fig. 4(c). This distance was merely limited by the current setup and should be reduced to a sub-millimeter scale in a practical fully packaged module, which can be comparable to current IMDD receiver modules.

   A CW light at a wavelength of 1550 nm was generated from a tunable laser source (TLS) and split into two ports, which served as the signal and the pilot tone ports. At the signal port, a LiNbO$_3$ IQ modulator was used to generate a high-speed coherent optical signal. The Nyquist filter was applied to the driving electrical signals from an arbitrary waveform generator (AWG). The modulated optical signal was then combined with the pilot tone by a polarization beam combiner (PBC). The optical power of the pilot tone was adjusted by a variable optical attenuator (VOA), so that their powers were nearly balanced. The self-coherent signal was then transmitted over a 25-km SMF. At the receiver side, the optical signal-to-noise ratio (OSNR) was controlled using another VOA, followed by an erbium-doped fiber amplifier (EDFA) and an optical bandpass filter (OBPF). The electrical signals from the six PDs of the PDA were amplified by differential RF amplifiers and then captured by a real-time oscilloscope (OSC). At a baudrate beyond 20 GBd, we could not use the balanced PD (B-PD) configuration due to the residual skew inside the PDA module. In these cases, we employed four single-ended PDs (S-PDs), where the electrical signals from $PD_x$, $PD_y$, $PD_a$, and $PD_l$ were independently captured by a four-channel real-time oscilloscope, so that the skew could be calibrated during DSP. By comparing the results using two configurations, the use of four S-PDs was validated (see Section S3 of Supplement 1 for details). To equalize and reconstruct the original IQ signal, we employed offline DSP with the 2×3 and 2×4 real-valued multi-input-multi-output (MIMO) equalizers [43, 44] for three-B-PD and four-S-PD configurations, respectively.

Figures 5(a)-(c) show the BER curves and the constellations for 15-GBd 16QAM signals, measured using the three-B-PD configuration. We can confirm that BERs well below the hard-decision forward error correction (HD-FEC) threshold are obtained with a negligible penalty even after 25-km transmission. Figures 5(d)-(f) show the results for 20-GBd 16QAM and 50-GBd QPSK signals, measured by the four-S-PD configuration. Once again, BERs below the HD-FEC threshold are obtained. Finally, Fig. 6 shows the measured BER curves and constellation diagrams of 15-GBd 16QAM signal at 1540-nm and 1565-nm wavelengths, demonstrating the wideband operation of our designed metasurface. While the baudrate in this work was limited by the bandwidth of the 2D-PDA, beyond-100-GBd transmission should be possible by using higher-speed surface-normal PDs with bandwidth exceeding 50 GHz [45, 46].

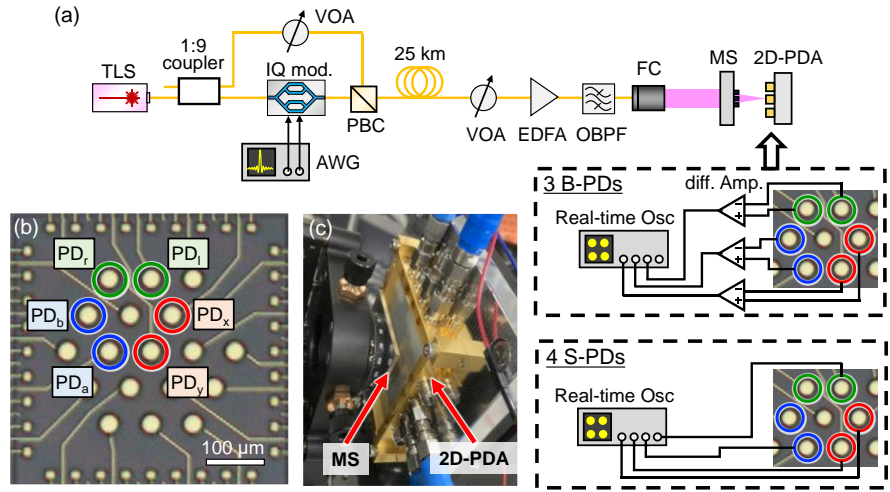

Fig. 4. Self-coherent transmission experiment using the fabricated metasurface and 2D-PDA. (a) Experimental setup. AWG: arbitrary waveform generator. PBC: polarization beam combiner. EDFA: erbium-doped fiber amplifier. OBPF: optical bandpass filter. Osc: oscilloscope. In the insets, three-B-PD and four-S-PD configurations are depicted. (b) Optical microscope image of the fabricated 19-pixel 2D-PDA. The six circled PDs were used in this experiment. (c) Photograph of the receiver.

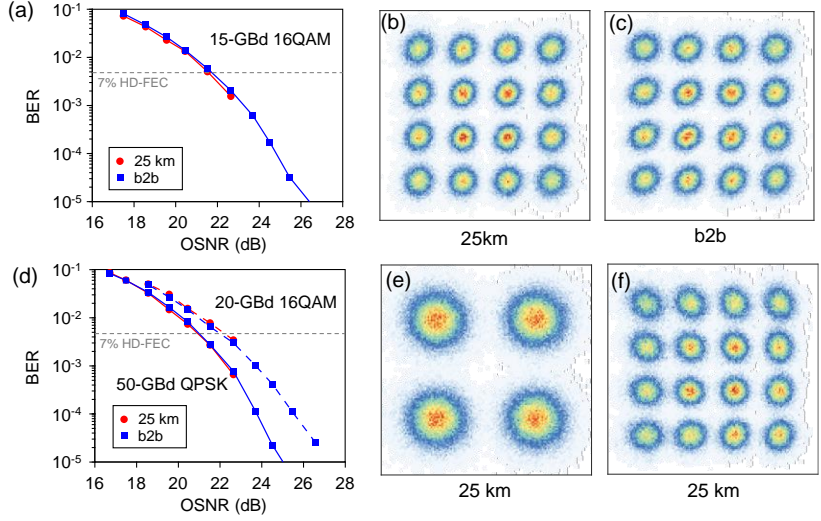

Fig. 5. Experimental results of self-coherent signal transmission at a wavelength of 1550 nm. (a)-(c) Measured BER curves and constellation diagrams of 15-GBd 16QAM signals before (b2b) and after 25-km transmission using the three-B-PD configuration. (d)-(f) Measured BER curves and constellation diagrams of 20-GBd 16QAM and 50-GBd QPSK signals after 25-km transmission using the four-S-PD configuration.

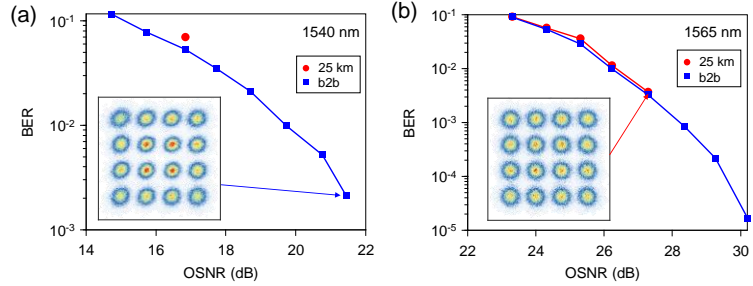

Fig. 6. Experimental results of self-coherent signal transmission at wavelengths of (a) 1540 nm and (b) 1565 nm. (a, b) Measured BER curves of 15-GBd 16QAM signals before (b2b) and after 25-km transmission using the three-B-PD configuration. The insets represent the retrieved constellation diagrams.

## 6. Conclusion

We have proposed and demonstrated a surface-normal SVR using a dielectric metasurface and 2D-PDA for the high-speed polarimetric self-coherent systems. Three independently designed meta-atom arrays based on Si nanoposts were superimposed onto a single thin metasurface layer to implement both the polarization-sorting and focusing functions simultaneously. Using a compact metasurface chip fabricated on a SOQ substrate, we demonstrated 25-km transmission of 20-GBd 16QAM and 50-GBd QPSK self-coherent signals. The operating baudrate was merely limited by the 2D-PDA, so that higher-capacity transmission should be possible by using a PDA with a broader bandwidth. Owing to the unique surface-normal configuration with the embedded lens array functionality, a compact receiver module with comparable size and complexity as the conventional IMDD receivers can be realized. Moreover, it can easily be extended to receive spatially multiplexed channels by simply replacing the SMF to a MCF and employing a larger-scale integrated PDA technology [32]. This work would, therefore, pave the way toward realizing cost-effective receivers for the future >Tb/s spatially multiplexed optical interconnects.

**Funding.** National Institute of Information and Communications Technology (NICT).

**Acknowledgments.** This work was obtained from the commissioned research 03601 by National Institute of Information and Communications Technology (NICT), Japan. Portions of this work were presented at the Optical Fiber Communications Conference (OFC) in 2022, M4J.5. A part of the device fabrication was conducted at the cleanroom facilities of d.lab in the University of Tokyo, supported by MEXT Nanotechnology Platform, Japan. The authors also thank all the technical staff at Advanced ICT device laboratory in NICT for supporting the PDA device fabrication. G.S. acknowledges the financial support from Optics and Advanced Laser Science by Innovative Funds for Students (OASIS) and World-leading Innovative Graduate Study Program - Quantum Science and Technology Fellowship Program (WINGS-QSTEP).

**References**

1. Cisco annual internet report (2018-2023), Cisco white paper. (2020).
2. W. Shieh and H. Ji, "Advanced direct detection schemes," in *Proc. Opt. Fiber Commun. Conf. (OFC) 2021*, paper Th4D.5.
3. D. Che, A. Li, X. Chen, Q. Hu, Y. Wang, and W. Shieh, "Stokes vector direct detection for linear complex optical channels," J. Lightwave Technol. **33**, 678–684 (2015).
4. T. Gui, X. Wang, M. Tang, Y. Yu, Y. Lu, and L. Li, "Real-time demonstration of homodyne coherent bidirectional transmission for next-generation data center interconnects," J. Lightwave Technol. **39**, 1231–1238 (2021).
5. T. Gui, H. Du, K. Zheng, J. Cao, S. Yuan, C. Yang, M. Tang, and L. Li, "Real time 6.4 Tbps (8×800G) SHCD transmission through 1+8 multicore fiber for co-packaged optical-IO switch applications," in *Proc. Opt. Fiber Commun. Conf. (OFC) 2022*, paper Th4C.1.
6. M. Mazur, A. Lorences-Riesgo, J. Schröder, P. A. Andrekson, and M. Karlsson, "High spectral efficiency PM-128QAM comb-based superchannel transmission enabled by a single shared optical pilot tone," J. Lightwave Technol. **36**, 1318–1325 (2018).
7. S. Ishimura, Y. Nakano, and T. Tanemura, "Impact of laser phase noise on self-coherent transceivers employing high-order QAM formats," J. Lightwave Technol. **39**, 6150–6158 (2021).
8. R. S. Luís, B. J. Puttnam, G. Rademacher, S. Shinada, and N. Wada, "Impact of GVD on polarization-insensitive self-homodyne detection receiver," IEEE Photonics Technol. Lett. **29**, 631–634 (2017).
9. K. Kikuchi, and S. Kawakami, "Multi-level signaling in the Stokes space and its application to large-capacity optical communications," Opt. Express **22**, 7374 (2014).
10. D. Che, A. Li, X. Chen, Q. Hu, Y. Wang, and W. Shieh, "Stokes vector direct detection for short-reach optical communication," Opt. Lett. **39**, 3110 (2014).
11. T. Hoang, M. Sowailem, Q. Zhuge, M. Osman, A. Samani, C. Paquet, S. Paquet, I. Woods, and D. Plant, "Enabling high-capacity long-reach direct detection transmission with QAM-PAM Stokes vector modulation," J. Lightwave Technol. **36**, 460-467 (2018).
12. H. Ji, D. Che, C. Sun, J. Fang, G. Milione, R. R. Unnithan, and W. Shieh, "High-dimensional Stokes vector direct detection over few-mode fibers," Opt. Lett. **44**, 2065 (2019).
13. P. Dong, X. Chen, K. Kim, S. Chandrasekhar, Y.-K. Chen, and J. H. Sinsky, "128-Gb/s 100-km transmission with direct detection using silicon photonic Stokes vector receiver and I/Q modulator," Opt. Express **24**, 14208–14214 (2016).
14. S. Ishimura, T. Fukui, R. Tanomura, G. Soma, Y. Nakano, and T. Tanemura, "64-QAM self-coherent transmission using symmetric silicon photonic Stokes-vector receiver," in *Proc. Opt. Fiber Commun. Conf. (OFC) 2022*, paper M4J.6.
15. S. Ghosh, T. Tanemura, Y. Kawabata, K. Katoh, K. Kikuchi, and Y. Nakano, "Decoding of multi-level Stokes-vector modulated signal by polarization-analyzing circuit on InP," J. Lightwave Technol. **36**, 187-194 (2018).
16. T. Suganuma, S. Ghosh, M. Kazi, R. Kobayashi, Y. Nakano, and T. Tanemura, "Monolithic InP Stokes vector receiver with multiple-quantum-well photodetectors," J. Lightwave Technol. **36**, 1268-1274 (2018).
17. S. Ghosh, T. Suganuma, S. Ishimura, Y. Nakano, and T. Tanemura, "Complete retrieval of multi-level Stokes vector signal by an InP-based photonic integrated circuit," Opt. Express **27**, 36449–36458 (2019).
18. M. Baier, F. M. Soares, A. Schoenau, Y. D. Gupta, D. Melzer, M. Moehrle, and M. Schell, "Fully integrated Stokes vector receiver for 400 Gbit/s," in *Proc. Opt. Fiber Commun. Conf. (OFC) 2019*, paper Tu3E.2.
19. T. Lei, C. Zhou, D. Wang, Z. Xie, B. Cai, S. Gao, Y. Xie, L. Du, Z. Li, A. V. Zayats, and X. Yuan, "On-chip high-speed coherent optical signal detection based on photonic spin-hall effect," Laser Photon. Rev. **16**, 2100669 (2022).
20. C. Zhou, Y. Xie, J. Ren, Z. Wei, L. Du, Q. Zhang, Z. Xie, B. Liu, T. Lei, and X. Yuan, "Spin separation based on-chip optical polarimeter via inverse design," Nanophotonics **11**, 813–819 (2022).
21. Y. Xie, T. Lei, D. Wang, J. Ren, Y. Dai, Y. Chen, L. Du, B. Liu, Z. Li, and X. Yuan, "High-speed Stokes vector receiver enabled by a spin-dependent optical grating," Photon. Res. **9**, 1470–1476 (2021).

22. A. Arbabi, Y. Horie, M. Bagheri, and A. Faraon, "Dielectric metasurfaces for complete control of phase and polarization with subwavelength spatial resolution and high transmission," Nat. Nanotechnol. **10**, 937–943 (2015).
23. E. Arbabi, S. M. Kamali, A. Arbabi, and A. Faraon, "Full-stokes imaging polarimetry using dielectric metasurfaces," ACS Photonics **5**, 3132–3140 (2018).
24. J. P. B. Mueller, K. Leosson, and F. Capasso, "Ultracompact metasurface in-line polarimeter," Optica **3**, 42–47 (2016).
25. A. Pors, M. G. Nielsen, and S. I. Bozhevolnyi, "Plasmonic metagratings for simultaneous determination of Stokes parameters," Optica **2**, 716–723 (2015).
26. Z. Yang, Z. Wang, Y. Wang, X. Feng, M. Zhao, Z. Wan, L. Zhu, J. Liu, Y. Huang, J. Xia, and M. Wegener, "Generalized Hartmann-Shack array of dielectric metalens sub-arrays for polarimetric beam profiling," Nat. Commun. **9**, 4607 (2018).
27. L. Li, J. Wang, L. Kang, W. Liu, L. Yu, B. Zheng, M. L. Brongersma, D. H. Werner, S. Lan, Y. Shi, Y. Xu, and X. Wang, "Monolithic full-Stokes near-infrared polarimetry with chiral plasmonic metasurface integrated graphene-silicon photodetector," ACS Nano **14**, 16634–16642 (2020).
28. P. J. Winzer and D. T. Neilson, "From scaling disparities to integrated parallelism: A decathlon for a decade," J. Lightwave Technol. **35**, 1099–1115 (2017).
29. D. A. B. Miller, "Attojoule optoelectronics for low-energy information processing and communications," J. Lightwave Technol. **35**, 346–396 (2017).
30. U. Koch, A. Messner, C. Hoessbacher, W. Heni, A. Josten, B. Baeuerle, M. Ayata, Y. Fedoryshyn, D. L. Elder, L. R. Dalton, and J. Leuthold, "Ultra-compact terabit plasmonic modulator array," J. Lightwave Technol. **37**, 1484-1491 (2019).
31. G. Soma, W. Yanwachirakul, T. Miyazaki, E. Kato, B. Onodera, R. Tanomura, T. Fukui, S. Ishimura, M. Sugiyama, Y. Nakano, and T. Tanemura, "Ultra-broadband surface-normal coherent optical receiver with nanometallic polarizers," ACS Photonics **9**, 2842–2849 (2022).
32. T. Umezawa, A. Matsumoto, K. Akahane, A. Kanno, and N. Yamamoto, "400-pixel high-speed photodetector for high optical alignment robustness FSO receiver," in *Proc. Opt. Fiber Commun. Conf. (OFC) 2022*, paper M4I.3.
33. E. Maguid, I. Yulevich, D. Veksler, V. Kleiner, M. L. Brongersma, and E. Hasman, "Photonic spin-controlled multifunctional shared-aperture antenna array," Science **352**, 1202–1206 (2016).
34. E. Arbabi, A. Arbabi, S. M. Kamali, Y. Horie, and A. Faraon, "Multiwavelength metasurfaces through spatial multiplexing," Sci. Rep. **6**, 32803 (2016).
35. T. Tanemura, T. Suganuma, and Y. Nakano, "Sensitivity analysis of photonic integrated direct-detection Stokes-vector receiver," J. Lightwave Technol. **38**, 447–456 (2020).
36. V. Liu and S. Fan, "S4: A free electromagnetic solver for layered periodic structures," Comput. Phys. Commun. **183**, 2233–2244 (2012).
37. J. P. Balthasar Mueller, N. A. Rubin, R. C. Devlin, B. Groever, and F. Capasso, "Metasurface polarization optics: Independent phase control of arbitrary orthogonal states of polarization," Phys. Rev. Lett. **118**, 113901 (2017).
38. M. Khorasaninejad, W. T. Chen, R. C. Devlin, J. Oh, A. Y. Zhu, and F. Capasso, "Metalenses at visible wavelengths: Diffraction-limited focusing and subwavelength resolution imaging," Science **352**, 1190–1194 (2016).
39. M. Khorasaninejad, A. Y. Zhu, C. Roques-Carmes, W. T. Chen, J. Oh, I. Mishra, R. C. Devlin, and F. Capasso, "Polarization-insensitive metalenses at visible wavelengths," Nano Lett. **16**, 7229–7234 (2016).
40. S. Molesky, Z. Lin, A. Y. Piggott, W. Jin, J. Vucković, and A. W. Rodriguez, "Inverse design in nanophotonics," Nat. Photonics **12**, 659–670 (2018).
41. M. Mansouree, A. McClung, S. Samudrala, and A. Arbabi, "Large-scale parametrized metasurface design using adjoint optimization," ACS Photonics **8**, 455–463 (2021).
42. T. Umezawa, T. Sakamoto, A. Kanno, S. Nakajima, A. Matsumoto, N. Yamamoto, and T. Kawanishi, "400-Gbps space division multiplexing optical wireless communication using two-dimensional photodetector array," in *Proc. Eur. Conf. Opt. Commun. (ECOC) 2018*, paper Th2.31.
43. Y. Mori, C. Zhang, and K. Kikuchi, "Novel configuration of finite-impulse-response filters tolerant to carrier-phase fluctuations in digital coherent optical receivers for higher-order quadrature amplitude modulation signals," Opt. Express **20**, 26236–26251 (2012).
44. M. S. Faruk and K. Kikuchi, "Compensation for in-phase/quadrature imbalance in coherent-receiver front end for optical quadrature amplitude modulation," IEEE Photonics J. **5**, 7800110 (2013).
45. Y. Yi, T. Umezawa, A. Kanno, and T. Kawanishi, "50 GHz high photocurrent PIN-PD and its thermal effect," in *Proc. OptoElectron. and Commun. Conf. (OECC) and Int. Conf. on Photon. in Switching and Computing (PSC) 2022*, paper WD2-2.
46. T. Umezawa, A. Kanno, K. Kashima, A. Matsumoto, K. Akahane, N. Yamamoto, and T. Kawanishi, "Bias-free operational UTC-PD above 110 GHz and its application to high baud rate fixed-fiber communication and W-band photonic wireless communication," J. Lightwave Technol. **34**, 3138–3147 (2016).

# Compact and scalable polarimetric self-coherent receiver using dielectric metasurface: Supplementary Information

Go Soma,[1,4] Yoshiro Nomoto,[2] Toshimasa Umezawa,[3] Yuki Yoshida,[3] Yoshiaki Nakano,[1] and Takuo Tanemura[1,5]

[1]*School of Engineering, The University of Tokyo, 7-3-1 Hongo, Bunkyo-ku, Tokyo, Japan*
[2]*Central Research Laboratory, Hamamatsu Photonics K.K. 5000 Hirakuchi, Hamakita-ku, Hamamatsu City, Shizuoka, Japan*
[3]*National Institute of Information and Communications Technologies (NICT), 4-2-1 Nukui-Kitamachi, Koganei, Tokyo, Japan*
[4]*soma@hotaka.t.u-tokyo.ac.jp*, [5]*tanemura@ee.t.u-tokyo.ac.jp*

## S1. Derivation of $D_u(\varphi_u, \varphi_v)$ and $D_v(\varphi_u, \varphi_v)$

From RCWA simulation, we obtain $t_u(D_u, D_v)$ and $t_v(D_u, D_v)$, which describe the complex transmittance through an array of elliptical Si nanoposts with the principal axis lengths of $D_u$ and $D_v$ for the linearly polarized light along $u$ and $v$ axes, respectively. The simulated intensity $|t_u|^2$ and $|t_v|^2$ and the phase $\arg(t_u)$ and $\arg(t_v)$ are shown in Fig. S1(a) and (b). From these results, we derive the required $D_u(\varphi_u, \varphi_v)$ and $D_v(\varphi_u, \varphi_v)$ to obtain desired phase shifts $(\varphi_u, \varphi_v)$, by using the following the equation [1]:

$$\left(D_u(\varphi_u, \varphi_v), D_v(\varphi_u, \varphi_v)\right) = \arg\min_{(D_u, D_v)} \left[\left|t_u(D_u, D_v) - e^{i\varphi_u}\right|^2 + \left|t_v(D_u, D_v) - e^{i\varphi_v}\right|^2\right].$$

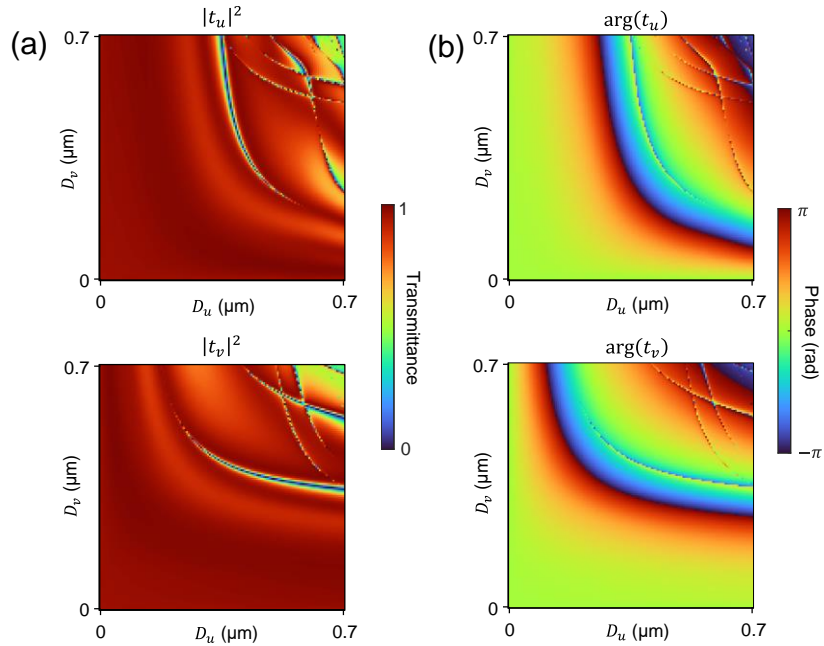

Fig. S1. Simulated intensity (a) and phase (b) of transmission coefficients as a function of $D_u$ and $D_v$.

## S2. Derivation of phase shift by a meta-atom array for circularly polarized light

The Jones matrix, describing the transmittance through a lossless Si nanopost array can be written as

$$\mathbf{J} = \mathbf{R}(\theta) \begin{pmatrix} e^{i\varphi_u} & 0 \\ 0 & e^{i\varphi_v} \end{pmatrix} \mathbf{R}(-\theta) = \begin{pmatrix} e^{i\varphi_u} \cos^2\theta + e^{i\varphi_v} \sin^2\theta & (e^{i\varphi_u} - e^{i\varphi_v})\sin\theta\cos\theta \\ (e^{i\varphi_u} - e^{i\varphi_v})\sin\theta\cos\theta & e^{i\varphi_u} \sin^2\theta + e^{i\varphi_v} \cos^2\theta \end{pmatrix}.$$

Here, $\varphi_u$ and $\varphi_v$ represent the phase shifts for the polarization components along the principal axes of the elliptical nanoposts and $\mathbf{R}(\theta)$ is a rotation matrix with a rotation angle of $\theta$. Here we assume that the input lightwave is circularly-polarized and its Jones vector is written as $\mathbf{E}_{r,l} = 1/\sqrt{2}(1, \pm i)^T$. Then, the output Jones vector is written as

$$\mathbf{J}\mathbf{E}_{r,l} = \frac{e^{i\varphi_u} + e^{i\varphi_v}}{2} \mathbf{E}_{r,l} + \frac{e^{i\varphi_u} - e^{i\varphi_v}}{2} e^{\pm i2\theta} \mathbf{E}_{l,r}.$$

Therefore, when the meta-atom functions as a half-wave plate, i.e., $\varphi_v = \varphi_u + \pi$, the output Jones vector becomes $e^{i(\varphi_u \pm 2\theta)} \mathbf{E}_{l,r}$. In other words, the phase shifts given to the right-handed and left-handed circularly-polarized waves are $(\varphi_r, \varphi_l) = (\varphi_u + 2\theta, \varphi_u - 2\theta)$, while the output polarization handedness is reversed.

## S3. Comparison between three-B-PD and four-S-PD configurations

To compare the results obtained by three-B-PD and four-S-PD configurations, we performed the self-coherent transmission experiment with 15-GBd 16QAM signals using the two experimental setups shown in Fig. 4(a). The measured BER curves are shown in Fig. S2. Since the BER was limited by the optical signal-to-noise ratio (OSNR), identical results were obtained in two cases.

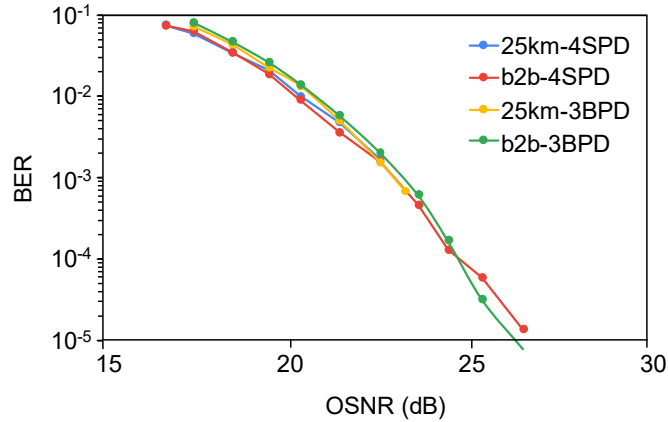

Fig. S2. Measured BER curves of 15-GBd 16QAM signals using the setup with three-B-PD and four-S-PD configurations.

## References

1. A. Arbabi, Y. Horie, M. Bagheri, and A. Faraon, "Dielectric metasurfaces for complete control of phase and polarization with subwavelength spatial resolution and high transmission," Nat. Nanotechnol. **10**, 937–943 (2015).